\documentclass[aps, prl, reprint, superscriptaddress,  floatfix]{revtex4-1}
\usepackage{latexsym}
\usepackage{amsthm}
\usepackage{amssymb}
\usepackage{amsmath}
\usepackage[all]{xy}
\usepackage{float}
\usepackage{graphicx}
\usepackage{subfigure}
\usepackage{dcolumn}
\usepackage{bm}
\usepackage{bbm}
\usepackage[usenames, dvipsnames]{xcolor}
\usepackage{amsfonts}
\usepackage{cases}
\usepackage{multirow}
\usepackage{bigstrut}
\usepackage[pagebackref=false, colorlinks=true,
linkcolor=blue, citecolor=blue,urlcolor=blue,]{hyperref}

\begin{document}

\title{Measurement of the earth tides with a diamagnetic-levitated micro-oscillator at room temperature}

\author{Yingchun Leng}
\affiliation{National Laboratory of Solid State Microstructures and Department of Physics,  Nanjing University, Nanjing 210093,
	China}

\author{Yiming Chen}
\affiliation{National Laboratory of Solid State Microstructures and Department of Physics, Nanjing University, Nanjing 210093,
	China}

\author{Rui Li}
\affiliation{CAS Key Laboratory of Microscale Magnetic Resonance and School of Physical Sciences, University of Science and Technology of China, Hefei 230026, China}

\affiliation{CAS Center for Excellence in Quantum Information and Quantum Physics, University of Science and Technology of China, Hefei 230026, China}

\affiliation{Hefei National Laboratory, Hefei 230088, China}

\author{Lihua Wang}
\affiliation{National Laboratory of Solid State Microstructures and Department of Physics, Nanjing University, Nanjing 210093,
	China}

\author{Hao Wang}
\affiliation{National Laboratory of Solid State Microstructures and Department of Physics, Nanjing University, Nanjing 210093,
	China}

\author{Lei Wang}
\affiliation{National Laboratory of Solid State Microstructures and Department of Physics, Nanjing University, Nanjing 210093,
	China}

\author{Han Xie}
\affiliation{National Laboratory of Solid State Microstructures and Department of Physics, Nanjing University, Nanjing 210093,
	China}

\author{Chang-Kui Duan}
\affiliation{CAS Key Laboratory of Microscale Magnetic Resonance and School of Physical Sciences, University of Science and Technology of China, Hefei 230026, China}

\affiliation{CAS Center for Excellence in Quantum Information and Quantum Physics, University of Science and Technology of China, Hefei 230026, China}

\affiliation{Hefei National Laboratory, Hefei 230088, China}

\author{Pu Huang}
\email{hp@nju.edu.cn}
\affiliation{National Laboratory of Solid State Microstructures and Department of Physics, Nanjing University, Nanjing 210093, China}

\author{Jiangfeng Du}
\email{djf@ustc.edu.cn}
\affiliation{CAS Key Laboratory of Microscale Magnetic Resonance and School of Physical Sciences, University of Science and Technology of China, Hefei 230026, China}
\affiliation{CAS Center for Excellence in Quantum Information and Quantum Physics, University of Science and Technology of China, Hefei 230026, China}
\affiliation{Hefei National Laboratory, Hefei 230088, China}
\affiliation{Institute of Quantum Sensing and School of Physics, Zhejiang University, Hangzhou 310027, China}

\begin{abstract}
The precise measurement of the gravity of the earth plays a pivotal role in various fundamental research and application fields.
Although a few gravimeters have been reported to achieve this goal, miniaturization of high-precision gravimetry remains a challenge.
In this work, we have proposed and demonstrated a miniaturized gravimetry operating at room temperature based on a diamagnetic levitated micro-oscillator with a proof mass of only $\rm 215~mg$.
Compared with the latest reported miniaturized gravimeters based on Micro-Electro-Mechanical Systems, the performance of our gravimetry has substantial improvements in that an acceleration sensitivity of $15~{\rm \mu Gal/\sqrt{Hz}}$ and a drift as low as $61~\rm \mu Gal$ per day have been reached.
Based on this diamagnetic levitation gravimetry, we observed the earth tides, and the correlation coefficient between the experimental data and theoretical data reached $0.97$.
Some moderate foreseeable improvements can develop this diamagnetic levitation gravimetry into chip size device, making it suitable for mobile platforms such as drones. Our advancement in gravimetry is expected to facilitate a multitude of applications, including underground density surveying and the forecasting of natural hazards.

\end{abstract}

\maketitle
\textit{Introduction.}\textbf{---}Starting from Galileo, the accurate measurement of the earth's gravity has been an important topic in physics. Measuring the earth's gravity, or the changes in the gravitational field, is not only a key technology in fundamental research, such as testing the equivalence principle, isotropy of post-Newtonian gravity and determining the Newtonian gravitational constant\cite{PhysRevLett.125.191101,PhysRevLett.100.031101,Rosi_Sorrentino_Cacciapuoti_Prevedelli_Tino_2014}, but also crucial in many application fields, such as predicting volcanic eruptions, monitoring earthquakes, and detecting underwater and underground resources or targets\cite{Antoni-Micollier2022,Montagner_Juhel2016,Stray_Lamb2022}. Gravimeters can be divided into two categories: absolute gravimeters and relative gravimeters. Absolute gravimeters acquire the gravitational acceleration $g$ by directly measuring a free-falling object, including the classic absolute gravimeter\cite{VanCamp_Williams_Francis_2005}, which uses a prism as the free-falling object, and quantum absolute gravimeter\cite{Kasevich_Chu_1992}, which use a group of atoms. While absolute gravimeters offer the advantage of minimal drift, they are bulky and expensive.

Relative gravimeters measure the relative variations of local gravitational acceleration by detecting the displacement of an oscillator. Compared to absolute gravimeters, solid-state relative gravimeters can achieve significantly smaller sizes. Examples include the  fused-quartz gravimeter\cite{Onizawa_2019} and the recently developed MEMS gravimeters\cite{Middlemiss_Samarelli_Paul_Hough_Rowan_Hammond_2016,Tang_Liu2019}. However, their long-term deployment and accuracy are limited by a rather large drift caused by the stress relaxation effect inherent in the solid-state system. Especially, for low-frequency geodynamical signals measurements, where gravity changes occur over time scales of days to years, such as hydrology, glacial isostatic adjustment, and Chandler wobble, low drift of the gravimeter is necessary\cite{surfacegravity}. Another type of relative gravimeter is the superconducting gravimeter\cite{10.1063/1.1683645}, which belongs to levitated mechanical system and has a very low drift. However,  superconducting gravimeters need to operate at extremely low temperatures to maintain superconductivity, which makes them bulky and expensive like absolute gravimeters. Therefore, there is a need to develop a lightweight, low-drift relative gravimeter that operates at room temperature.

Levitated mechanical systems have been developed in recent years based on different methods, including superconducting magnetic levitation\cite{PhysRevLett.131.043603,10.1063/1.5129145}, diamagnetic levitation\cite{Slezak_2018},
electric levitation\cite{PhysRevResearch.2.023349},
optical levitation\cite{Gieseler_Novotny_Quidant_2013,Ahn_Xu_Bang_Ju_Gao_Li_2020}, and so on. Thanks to the ultrahigh force detection sensitivity of these levitated mechanical oscillators, they have been widely used in the field of detecting  weak signals due to charge, spin, temperature, etc.\cite {PhysRevLett.113.251801,Neukirch_von2015,Millen_Deesuwan_Barker_Anders_2014},
and have also been applied in the exploration of fundamental physics like the dark energy model\cite{Yin_Li_Yin_2022}, wave function collapse models\cite{PhysRevResearch.2.023349,PhysRevResearch.2.013057}, and observation of quantum
effects\cite{PhysRevLett.124.163604,doi:10.1126/science.aba3993}. Among these levitation methods, the diamagnetic levitation and superconducting levitation have been achieved with resonant frequency as low as a few hertz, making them ideal for detecting weak gravity changes. Recently, acceleration sensitivities lower than 10 $\mu \rm Gal/\sqrt{Hz}$ have been achieved with diamagnetic levitation oscillators (1 $\rm Gal$ = 1 $\rm cm/s^2$)\cite{Yin_Li_Yin_2022,PhysRevApplied.16.L011003}, demonstrating the ability to detect weak acceleration variations (a few $\mu \rm Gal$) under the large earth gravitational acceleration background $g$ (980 $\rm Gal$). The diamagnetic-levitated micro-oscillator, due to its lack of mechanical losses inherent in solid-state systems and its small volume, offers a promising avenue to implement a miniaturized, low-drift relative gravimeter at room temperature.

In this work, we report the successful measurement of the earth tides using a diamagnetic levitation gravimeter at room temperature. The correlation coefficient between the experimental results and the theoretical calculation of the earth tides is $0.97$, the corresponding acceleration sensitivity is 15  $\mu \rm Gal/\sqrt{Hz}$ at 1 Hz, and a bias instability of 5.1 $\mu \rm Gal$ is observed. More importantly, the drift of this method is only 61 $\mu \rm Gal$ per day, one of the best values among relative gravimeters. Finally, we discuss the future development of miniaturized low-drift gravimeters using this approach and their potential  applications in underground resources detection and other areas.

\begin{figure}
	
	\includegraphics[width=0.92\columnwidth]{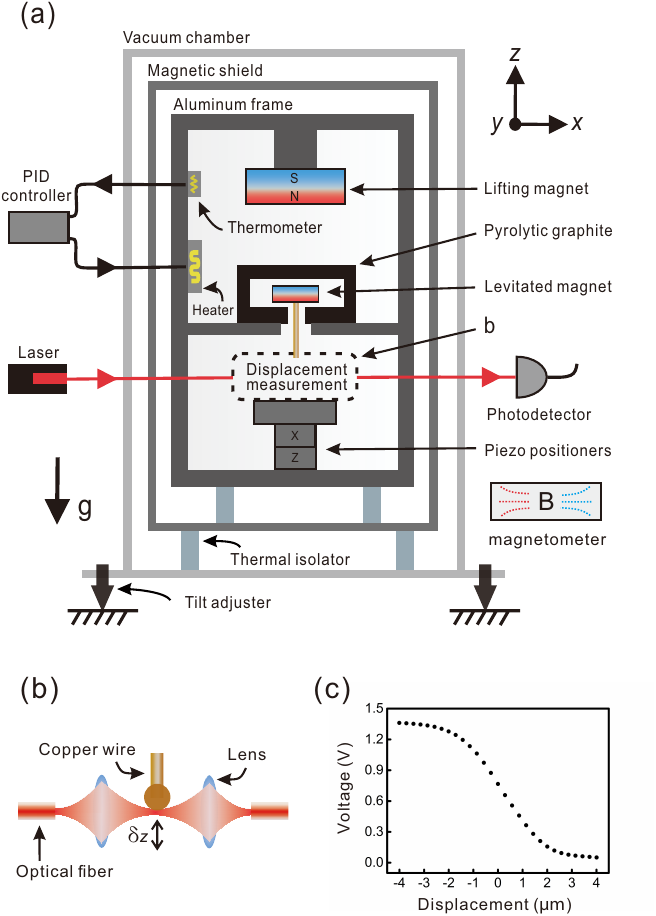}
	\caption{(Color online). Schematic of the diamagnetic levitation gravimeter. \textbf{a}, The diamagnetic levitated micro-oscillator with a proof mass of 215 mg is placed in the aluminum frame. The thermometer and heater are fixed on the aluminum frame to stabilize the temperature of the permanent magnets with the PID controller. A vacuum chamber and thermal isolators are used to avoid ambient temperature disturbances. Magnetic shields are placed outside the aluminum frame to shield against external magnetic field. A 633-nm laser beam reaches the displacement sensor and  is then sent to a photodetector through optical fibers. Piezoelectric positioners are used to control the position of displacement sensor to ensure that the copper wire is exactly at the focal point of the laser beam. Tilt adjusters are used to level the setup. A magnetometer is placed outside of the vacuum chamber to measure the external magnetic field. \textbf{b}, Schematic of oscillator displacement detection. The laser beam is focused by a lens. The copper wire is placed on the focal point, where displacement sensitivity in Z direction is maximum. \textbf{c}, The measured response curve of voltage to displacement in Z direction of \textbf{b}. It can be seen that the response curve has very good linearity in the range of -1 $\mu \rm m$ to 1 $\mu \rm m$, with the R-squared of linear fitting for this segment of the data reaching 0.999.}
	\label{exp-sys}\hypertarget{exp-sys}{}
\end{figure}

\textit{Experimental system.}\textbf{---}For a harmonic oscillator, the response of displacement $x(\omega)$ to acceleration $a(\omega)$ at frequency $\omega/2\pi $ is:
\begin{equation}
	\label{displacement} x(\omega)=\frac{a(\omega)}{\sqrt{(\omega_0^2-\omega^2)^2+(\gamma \omega)^2}}
\end{equation}
where $\omega_0/2\pi$ is the resonant frequency and $\gamma/2\pi $ is the dissipation of the oscillator. In the region far below resonance, where gravimeters operate, the displacement is inversely proportional to the square of the resonant frequency. Therefore, the response of displacement to the gravity signal can be improved by decreasing the resonant frequency in Z direction of the oscillator. The experimental device, as shown in
Fig.\ \hyperlink{exp-sys}{1(a)}, uses a levitated magnet placed right below the lifting magnet as an oscillator. At the equilibrium position, the lifting force produced by the lifting magnet is balanced by the gravity of the levitated magnet, resulting in a net force of zero. Two diamagnetic slabs made of pyrolytic graphite are placed above and below the levitated magnet to generate a low-frequency trap in Z direction\cite{10.1119/1.1375157}.  By tuning the gap spacing between two pyrolytic graphite slabs, the frequency of resonance mode along Z direction is tuned to as low as 1 Hz. Adding pyrolytic graphite slabs around the oscillator raises the resonant frequency of X and Y modes to suppress displacement in X and Y directions of the oscillator, overcoming disturbances caused by floor tilt.

Fig.\ \hyperlink{exp-sys}{1(b)} shows the schematic of oscillator displacement detection\cite{Yin_Li_Yin_2022}. The displacement of the oscillator is obtained by detecting the position of the copper wire that is glued to the surface of the levitated magnet. A laser beam with a wavelength of 633 nm enters the vacuum chamber through optical fibers and then is focused by a lens. The copper wire is placed at the focal point to achieve maximum displacement response along the Z direction. The displacement responses in X and Y directions are an order of magnitude lower than in the Z direction, which also helps reduce disturbances caused by floor tilt. Fig.\ \hyperlink{exp-sys}{1(c)} shows the measured response curve of voltage to displacement in Z direction for this detection device. The linear response range of voltage to displacement is 2 $\mu \rm m$. For an oscillator with resonant frequency of 1 Hz, this displacement range corresponds to acceleration variations of 7800 $\mu \rm Gal$, while the acceleration variations caused by the earth tides are less than 400 $\mu \rm Gal$. The detection device is placed on the X, Z piezoelectric positioners. By moving these positioners, the copper wire can be precisely positioned within this linear region to measure the displacement of the oscillator.

The levitated magnet is naturally subject to the torque and the force exerted by the magnetic field and its gradient. Hence, we have incorporated magnetic shields to mitigate unwanted external magnetic field. However, the displacement of oscillator due to the external magnetic field within the laboratory is still observable.
The temporal change of the spatial gradient of external magnetic field can be regarded as proportional to the temporal change in the external magnetic field, denoted as $\delta{B(t)}$. This is because the artificial external magnetic field source and the gravimeter can be treated as relatively static during the experiment. Therefore, the variation of the force on the oscillator caused by the spatial gradient of the external magnetic field is proportional to $\delta{B(t)}$. The variation in torque on the oscillator generated by the external magnetic field is also proportional to $\delta{B(t)}$. As $\delta{B(t)}$ in the laboratory is small and its frequency is much lower than the oscillator's resonant frequency, the displacement of oscillator in translation and rotation modes caused by an external magnetic field is proportional to $\delta{B(t)}$. A typic standard deviation of 0.86 $\mu \rm T$ for the external magnetic field has been observed, resulting an acceleration noise of 137  $\mu \rm Gal$. Therefore, a magnetometer has been used to measurement the change of external magnetic field to reduce this impact by correction \cite{SM}.

For permanent magnets, the remanence is temperature-dependent, causing  the position of the oscillator to change with the temperature of the diamagnetic levitated mechanical system. To reduce the temporal change of temperature, denoted as $ \delta T(t)$, we use SmCo magnets as the lifting and levitated magnets due to their low temperature coefficient of remanence. In addition, the entire diamagnetic levitated mechanical setup is placed in a vacuum chamber with a pressure of $10^{-5} $ mbar, and thermal isolators made of Peek are used as supporting legs of the magnetic shields and aluminum frame to minimize heat exchange between the magnets and external environment. A PID controller is also used on the aluminum frame to stabilize the temperature of the two magnets. A standard deviation of 20 $\mu \rm K$ for the temperature of the lifting magnet has been achieved, resulting in an acceleration noise of 5 $\mu \rm Gal$. The temperature is also recorded for reducing this effect \cite{SM}.

\begin{figure}
	\centering
	\includegraphics[width=1.0\columnwidth]{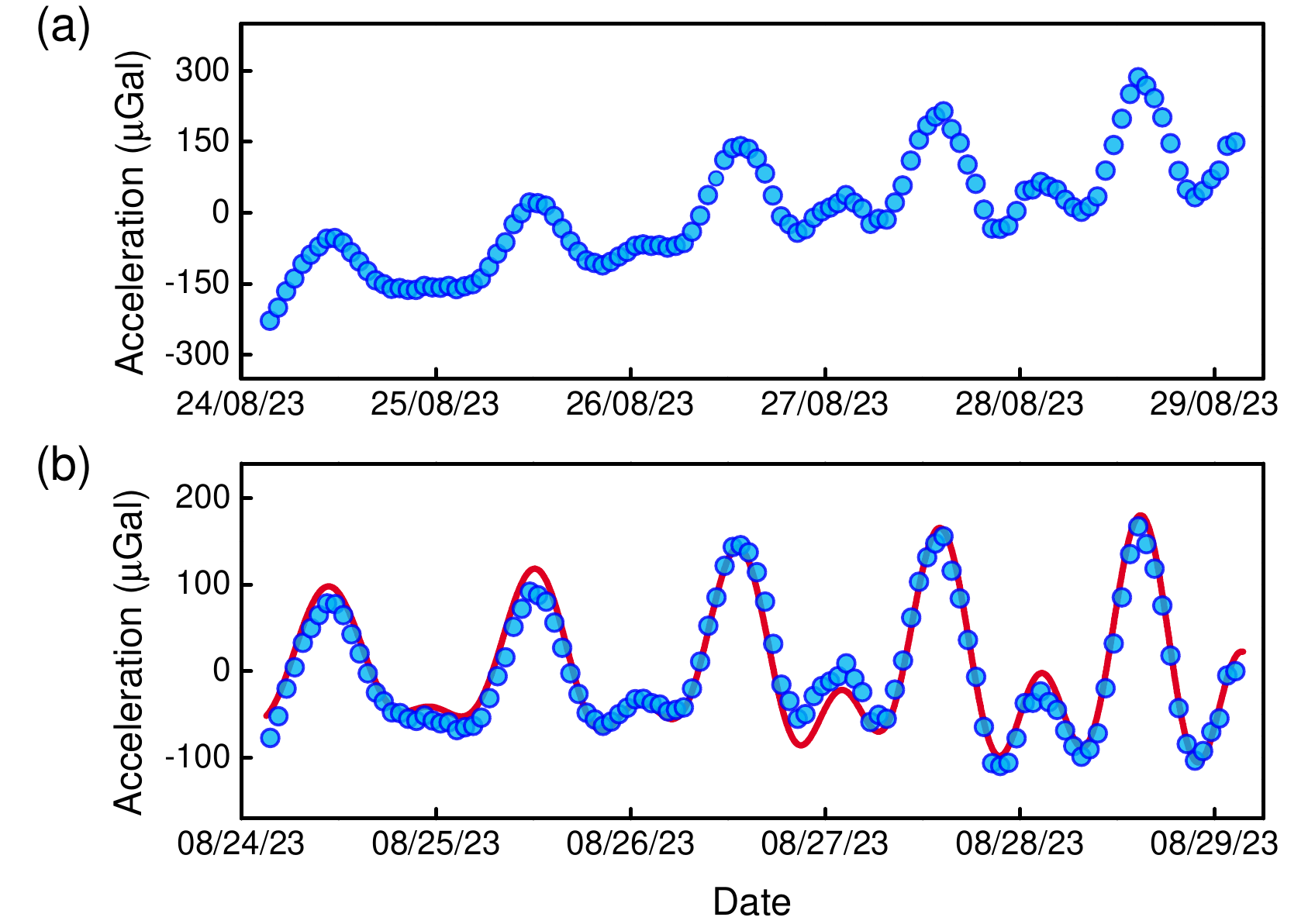}
	\caption{(Color online). Measurement of the earth tides. \textbf{a}, Data obtained from the diamagnetic levitated gravimeter. Each blue dot represents the mean value of measured data over a 1-hour period. A drift of 61 $\mu \rm Gal$ per day has been observed. \textbf{b}, Comparison between the theoretical and experimental data of the earth tides. The red line is a theoretical plot calculated with $TSOFT$\cite{VANCAMP2005631}. The blue dots represent the same data as in \textbf{a}, but with the drift removed. The correlation coefficient between experimental data and theoretical calculation is 0.97.}
	\label{results}\hypertarget{results}{}
\end{figure}

\begin{figure*}
	\includegraphics[width=2.0\columnwidth]{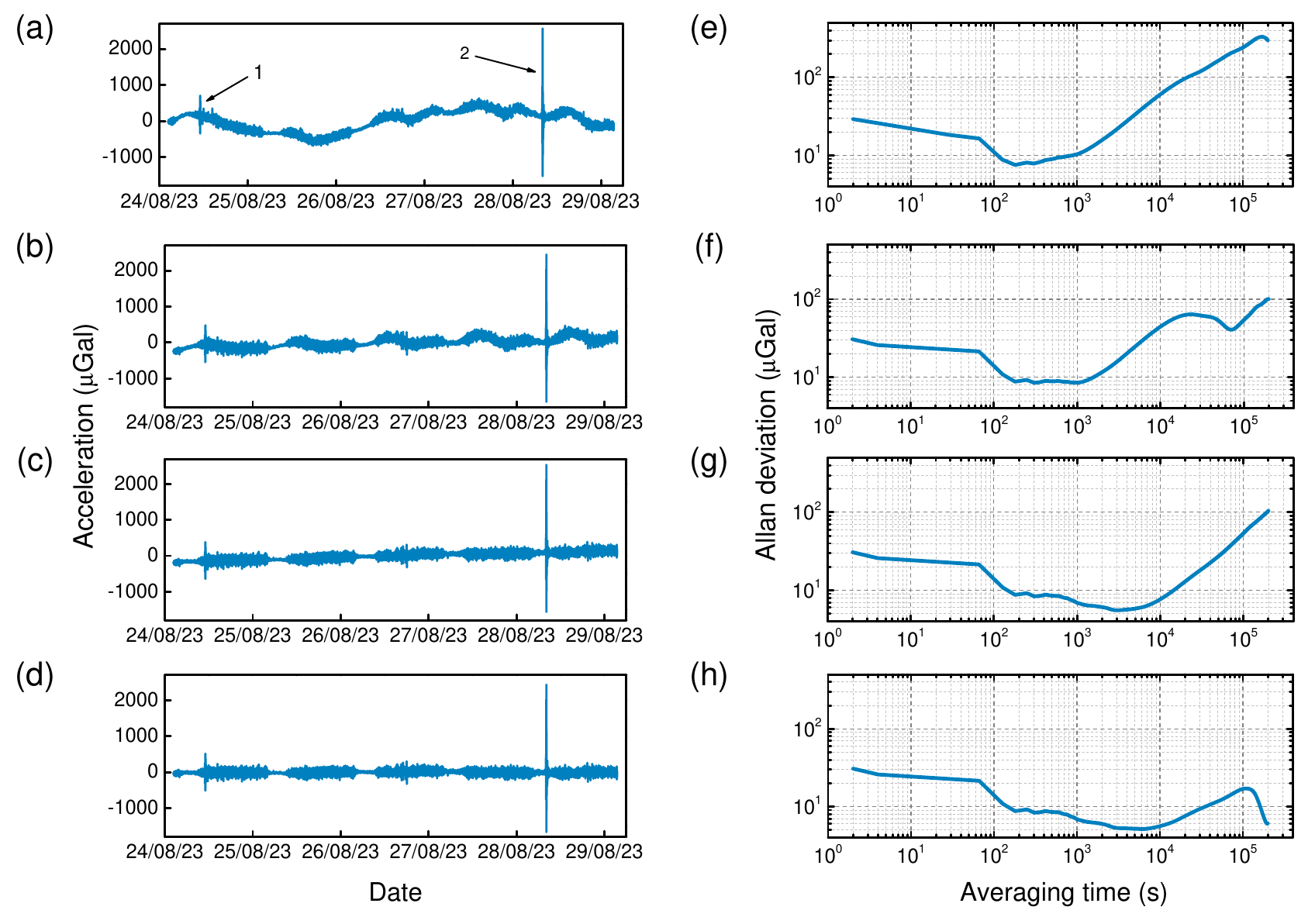}
	\caption{(Color online). Allan deviation. \textbf{a}, Raw data with a sampling rate of 0.5 Hz. Points 1 and 2 are earthquake signals that occurred off the east coast of Honshu, Japan and in the Bali Sea, Indonesia, respectively. \textbf{b}, Data from \textbf{a} corrected for external magnetic field and temperature. \textbf{c}, Data from \textbf{b} with the theoretical tide signal removed. \textbf{d}, Data from \textbf{c} with drift removed. \textbf{e}$-$\textbf{h}, represent the Allan deviation of the data in \textbf{a}$-$\textbf{d}, respectively. The bias instability of this gravimeter is 5.1 $\mu \rm Gal$ at an average time of 6000 s.}
	\label{Allan}\hypertarget{Allan}{}
\end{figure*}
\textit{The measurement of the earth tides.}\textbf{---}Local gravitational acceleration experiences small variations due to the earth tides, which result from changes in the relative phase between the Sun, Moon, and Earth. The magnitude of these variations depends on local latitude, altitude, and measurement time. the earth tides are a natural weak gravity signal that can be used to demonstrate the sensitivity and long-term stability of gravimeters. The experimental setup is placed directly on the floor of the basement without seismic isolation. After the vacuum chamber was pumped, the magnitude of the drift exponentially decayed over time for the first few weeks and it took several weeks for the temperature and drift of the experimental system to stabilize \cite{SM}, then the data were recorded continuously for 5 days from 24 to 29 August 2023 in the laboratory (32.059°N, 118.771°E). The data plotted in Fig.\ \hyperlink{results}{2(a)} represent experimentally measured acceleration variations over time, averaged with a time constant of 1 hour. From this, a drift of the system at 61 $\mu \rm Gal$ per day is observed, which is almost an order of magnitude lower than that of the fused-quartz gravimeter (500 $\mu \rm Gal$ per day) and twice as low as the Glasgow MEMS gravimeter (140  $\mu \rm Gal$ per day)\cite{Middlemiss_Samarelli_Paul_Hough_Rowan_Hammond_2016}. Fig.\ \hyperlink{results}{2(b)} compares the theoretical calculation of the earth tides using $TSOFT$ software\cite{VANCAMP2005631} and experimental measurements. The system's drift has been removed from the experimental data. The correlation coefficient between the experimental data and theoretical calculation reached 0.97. Using this earth tide signal to calibrate the gravimeter, we determined that the acceleration sensitivity of the device is 15 $\mu \rm Gal/\sqrt{Hz}$ \cite {SM}.

Allan deviation is a useful technique to investigate the stability of a device\cite{1446564}.
Fig.\ \hyperlink{Allan}{3} shows the experimental data in the time domain and the corresponding Allan deviation. Fig.\ \hyperlink{Allan}{3(a)} displays a full-noise measurement data set, with spikes marked with number 1 and 2 representing earthquake signals that occurred off the east coast of Honshu, Japan and in the Bali Sea, Indonesia, respectively. The noise measured in the early hours of the morning is noticeably smaller than during the day, which is consistent with the external magnetic field noise in the laboratory. Fig.\ \hyperlink{Allan}{3(b)} shows the same data as in Fig.\ \hyperlink{Allan}{3(a)}, but corrected for external magnetic field and temperature signals through a regression method. Floor tilt variations were also recorded during the experiment, but no  discernible effect on the raw data was observed. Fig.\ \hyperlink{Allan}{3(c)} presents the same data as in
Fig.\ \hyperlink{Allan}{3(b)}, but with the earth tide signal removed. Fig.\ \hyperlink{Allan}{3(d)} plots the same data as in
Fig.\ \hyperlink{Allan}{3(c)}, but with drift removed. Fig.\ \hyperlink{Allan}{3(e)}$-$\hyperlink{Allan}{3(h)} show the Allan deviation of data in Fig.\ \hyperlink{Allan}{3(a)}$-$\hyperlink{Allan}{3(d)}, respectively. The broad peak from $10^{4} $ s to $10^{5} $ s in Fig.\ \hyperlink{Allan}{3(f)} is the tide signal. A bias instability of 5.1 $\mu \rm Gal$ is obtained with an integration time of 6000 s, as shown in Fig.\ \hyperlink{Allan}{3(h)}.

\begin{table}[h]
	\caption{\label{tab:table1} Comparison of key parameters between this work and typical relative gravimeters operating at room temperature.}	
	\begin{ruledtabular}
		\renewcommand{\arraystretch}{1.25}
		\begin{tabular}{ccc}
		\multirow{2}{*}{Gravimeters} & Sensitivity                & Drift                         \\ 
		          	                 &  ($\mu \rm Gal/\sqrt{Hz}$)  & ($\mu \rm Gal/Day$)       \\	
			\hline
		Scintrex CG-5 \cite{Onizawa_2019}   & 2           & 500                   \\
		Glasgow MEMS  \cite{Middlemiss_Samarelli_Paul_Hough_Rowan_Hammond_2016} 
		                                    & 40          & 140                     \\
		HUST MEMS     \cite{Tang_Liu2019}   & 8  		  & 2400                      \\
		This Work       					& 15          & 61                     \\
                                                                		
	\end{tabular}
	\end{ruledtabular}
\end{table}

\textit{Discussion and summary.}\textbf{---}We have observed the earth tides using a diamagnetic levitation gravimeter, the correlation coefficient between experimental data and theoretical calculation is 0.97. TABLE~\ref{tab:table1} is a comparison between this work and typical relative gravimeters operating at room temperature. An acceleration sensitivity of 15 $\mu \rm Gal/\sqrt{Hz}$ and a drift of 61 $\mu \rm Gal$ per day has been achieved, suitable for underground density surveying and the forecasting of natural hazards.
To put the current acceleration sensitivity in perspective for density contrast imaging, an ore body with a residual density of 2000 $\rm kg/m^{3}$ and a size of $20\times 20\times 20$ $\rm m^{3}$ at a depth of 84 m can be detected in one second.

Since the thermal noise of the oscillator ($0.8~\mu \rm Gal/\sqrt{Hz}$) is much lower than the measured acceleration noise (15 $\mu \rm Gal/\sqrt{Hz}$), the present acceleration sensitivity of our device is mainly limited by seismic noise and external magnetic field noise. Hence, the gravimeter can be further miniaturized by reducing the proof mass of the oscillator to several hundred $\mu \rm g$ without deteriorating its current acceleration sensitivity. The low magnetic field gradient required is as low as the order of $10^{-1}$ $\rm T/m $, which can be provided by an integrated coil. With these anticipated improvements, it will be feasible to integrate this device into a chip, making it suitable for a wide range of mobile measurement platforms including drones.

Regarding to the drift, it is in the opposite direction to the earth's gravity, which cannot be explained by the magnetic flux decay of permanent magnets. A possible reason could be that water molecules from the air attach to the surface of the levitated magnet but are slowly released in the vacuum during the experiment, leading to a gradual reduction in the weight of the levitated magnet over time.
Nevertheless, the magnetic flux decay of permanent magnets is expected to be the main factor contributing to the drift of diamagnetic levitation gravimetry. For SmCo magnets in air, the flux decay corresponds to drift about $10~\mu \rm Gal$ per day\cite{1983ITM19.2050D}. In a vacuum condition, where magnets are protected from oxidation, an ultralow drift at sub-$\mu$Gal scale  could be expected.

\hspace*{\fill}
	
This work was supported by the National Natural Science Foundation of China (grants no.\ 12150011 no.\ T2388102, no.\ 12075116, and no.\ 11890702), the Primary Research $\&$ Development Plan of Jiangsu Province (grant no.\ BE2021004-2), the Chinese Academy of Sciences (grant no.\ XDC07000000), the Anhui Initiative in Quantum Information Technologies (grant no.\ AHY050000) and the Hefei Comprehensive National Science Center.

 Y. L., Y. C., R. L., L. W. and H. W. contributed equally to this work.

\bibliographystyle{apsrev4-1}
%

\end{document}